\documentstyle[aps,prb]{revtex}
\newcommand{\bfr}{{\bf r}}
\newcommand{\bfv}{{\bf v}}
\newcommand{\bfb}{{\bf B}}
\newcommand{\bfe}{{\bf E}}
\newcommand{\bfrr}{(\bfr)}
\newcommand{\bfrrp}{(\bfr')}
\newcommand{\bfrt}{(\bfr,t)}
\newcommand{\grn}{G(\bfr,\bfr')}
\newcommand{\dfunr}{\delta (\bfr-\bfr')}
\newcommand{\exb}{{\bfe\times\bfb}}
\newcommand{\bge}{\begin{equation}}
\newcommand{\ee}{\end{equation}}
\newcommand{\bgea}{\begin{eqnarray}}
\newcommand{\eea}{\end{eqnarray}}
\newcommand{\bgeas}{\begin{eqnarray*}}
\newcommand{\eeas}{\end{eqnarray*}}
\newcommand{\laplacian}{\nabla^2}
\newcommand{\dtr}{{d^2r\;}}
\newcommand{\dtrp}{{d^2r'\;}}
\newcommand{\smallhalf}{\mbox{{\footnotesize $\frac{1}{2}$}}}
\newcommand{\threehalves}{\mbox{{\footnotesize $\frac{3}{2}$}}}
\newcommand{\sevenninths}{\mbox{{\footnotesize $\frac{7}{9}$}}}
\def\grad{\mbox{\boldmath $\nabla$}}
\newtheorem{axiom}{Axiom}
\newtheorem{property}{Property}

\begin{document}

\title{
\begin{flushleft}
{\small BU-CCS-950501}\\
\end{flushleft}
Thermodynamic Description of the Relaxation of Two-Dimensional Euler
Turbulence Using Tsallis Statistics\footnote{
\small \baselineskip=11pt This work was supported in part by the Air
Force Office of Scientific Research and the Department of Energy,
Division of High Energy Physics.}
%\\(DRAFT COPY -- DO NOT CIRCULATE)
}
\author{
Bruce M. Boghosian\\
{\small \sl Center for Computational Science,} \\
{\small \sl Boston University,} \\
{\small \sl 3 Cummington Street, Boston, Massachusetts 02215, U.S.A.}
%\\[0.3cm]
}
\date{\today}
\maketitle

\begin{abstract}
Euler turbulence has been experimentally observed to relax to a
metaequilibrium state that does not maximize the Boltzmann entropy, but
rather seems to minimize enstrophy.  We show that a recent
generalization of thermodynamics and statistics due to Tsallis is
capable of explaining this phenomenon in a natural way.  The
maximization of the generalized entropy $S_{1/2}$ for this system leads
to precisely the same profiles predicted by the Restricted Minimum
Enstrophy theory of Huang and Driscoll.  This makes possible the
construction of a comprehensive thermodynamic description of Euler
turbulence.
\end{abstract}

\vspace{0.25truein}


%PREPRINT:
%\clearpage

\section{Introduction}

The kinetic and field equations for the drift motion of a pure electron
plasma column in a strong magnetic field are isomorphic to the equations
of motion of a two-dimensional Euler fluid~\cite{bib:iso}.  The density
of the plasma corresponds to the vorticity of the fluid, and the
electrostatic potential corresponds to the stream function.  This
observation has made it possible to use pure electron plasmas to study
Euler turbulence in the laboratory~\cite{bib:nnreview}.  Such
experiments have followed the relaxation of Euler turbulence through
several identifiable stages~\cite{bib:stages}: An initially hollow
vorticity profile develops a linear diochotron instability which
saturates with the creation of long-lived vortex patches.  These patches
move about for hundreds of diochotron periods, shedding filiaments, and
eventually mixing and inwardly transporting.  This process gives rise to
an axisymmetric metaequilibrium state, whose density decreases
monotonically with radius, which then persists for tens of thousands of
diochotron periods.  The eventual decay of this state is due only to
viscous and three-dimensional effects that destroy the idealization of
the two-dimensional Euler fluid~\cite{bib:iso,bib:huang1}.

The shape of the radial vorticity profile of the metaequilibrium state
is an interesting and fundamental problem.  One would expect that it
could be described by a variational principle, but the most natural
principle of this sort -- the maximization of the Boltzmann entropy
under the constraints of constant mass, energy, and angular momentum --
has been found to yield profiles that are substantially flatter than
those observed in experiments~\cite{bib:huang1,bib:huang2}.  On the
other hand, an alternative variational principle, in which the enstrophy
(the integral of the square of the vorticity) is minimized has been
found to yield results in excellent agreement with
experiment~\cite{bib:huang1,bib:huang2}.  To date, however, there has
existed no satisfactory theoretical explanation for this unusual
variational principle.

In this paper we show that the failure of the Boltzmann entropy to
predict the radial density profile of the metaequilibrium state can be
understood as but one example of a systemic breakdown of Boltzmann-Gibbs
statistics for systems with long-range interactions, long-time
memory~\cite{bib:breakdown}, or fractal space-time
structure~\cite{bib:fractal}.  Moreover, we show that a recent
generalization of statistics and thermodynamics due to
Tsallis~\cite{bib:tsallis1} is capable of explaining this phenomenon
much more naturally: The maximization of the Tsallis entropy $S_q$ for
this system, with $q=\smallhalf$, leads to precisely the same profiles
predicted by the Restricted Minimum Enstrophy (RME) theory of Huang and
Driscoll~\cite{bib:huang2}.  This observation makes it possible to
develop a consistent thermodynamic description of such systems, and to
associate this phenomenon with a wide body of research on generalized
statistics and thermodynamics.

The outline of this paper is as follows: In Section~\ref{s:nonneutral}
we describe the dynamical equations of the nonneutral plasma column (or,
equivalently, of the two-dimensional Euler fluid), cast them in
Hamiltonian format, and present the constants of the motion.  We also
review the experimental results for this system, and describe previous
attempts to explain the metaequilibrium density profile by a variational
principle.  In Section~\ref{s:genthermo} we describe Tsallis'
generalization of thermodynamics, and in Section~\ref{s:stellar} we
review the application of Tsallis' formalism to the problem of stellar
polytropes, which are static solutions to the Poisson-Vlasov equations.
This problem was first considered by Plastino and
Plastino~\cite{bib:pp1}, who applied Tsallis' methods to a linear energy
functional.  Noting that the energy functional of the Poisson-Vlasov
system is, strictly speaking, quadratic~\cite{bib:morrison}, we redo
this analysis.  Finally, in Section~\ref{s:turb} we return to the
problem of the metaequilibrium state of two-dimensional Euler
turbulence, and we show that Tsallis' generalized thermodynamics may be
used to explain the observed density profiles.

\section{Noneutral Plasma Profiles and Euler Turbulence}
\label{s:nonneutral}

\subsection{Dynamical Equations}

Strongly magnetized pure-electron plasmas in ``Penning traps'' with
cylindrical geometry and electrostatic axial confinement have been
studied for some time now~\cite{bib:nnreview}.  Such plasmas typically
have a gyrofrequency that is much greater than the bounce frequency,
which in turn is much greater than the drift frequency.  That being the
case, we can average over the gyro and bounce time scales, and describe
the system by the two-dimensional drift motion of guiding centers,
perpendicular to the magnetic field.

Since the magnetic field is uniform, the dominant drift mechanism is the
$\exb$ drift, given by
\[
{\bf v}_\exb = c\frac{\exb}{B^2},
\]
where $\bfb$ is the applied magnetic field, $\bfe$ is the
self-consistent electric field, and $c$ is the speed of light.  Since
this drift velocity is independent of particle thermal velocity, it is
possible to project out the velocity degrees of freedom in phase space,
and thereby write a Vlasov equation directly for the guiding-center
density $n\bfrt$,
\[
0 = \frac{\partial n\bfrt}{\partial t} +
    {\bf v}_\exb\cdot\grad n\bfrt.
\]
Writing $\bfe=-\grad\Phi$, and adopting dimensionless units with a
magnetic field of unit magnitude, this can be written
\bge
0 = \frac{\partial n\bfrt}{\partial t} +
    \hat{{\bf b}}\cdot
    \left[\grad\Phi\times\grad n\bfrt\right],
 \label{eq:vlasov}
\ee
where $\hat{{\bf b}}$ is a unit vector in the direction of the magnetic
field, and where we have used the vector ``triple-product'' identity.
The self-consistent electrostatic potential is then given by the Poisson
equation,
\[
 \laplacian\Phi\bfrt = 4\pi e n\bfrt,
\]
where $-e$ is the electronic charge.  (Henceforth, we set $e=1$.)

If we identify $n$ as the vorticity and $\Phi$ as the stream function,
we note that these equations are isomorphic to Euler's equations of
inviscid fluid dynamics in two dimensions.  Likewise, Dirichlet boundary
conditions, for which the wall is an equipotential, correspond to the
condition that the normal velocity of the Euler fluid vanishes at the
wall.  Indeed, simulations of pure-electron plasma columns provide an
important experimental tool for the study of two dimensional Euler
turbulence.  Henceforth, we interchangibly refer to the physical
embodiment of this dynamical system as a pure-electron plasma column, or
as two-dimensional Euler turbulence.

\subsection{Hamiltonian Structure}

In spite of the fact that we have projected from phase space to
configuration space, we note that the Vlasov equation,
Eq.~(\ref{eq:vlasov}), has a symplectic Hamiltonian form in two
dimensions.  Specifically, taking the magnetic field in the
$z$-direction, so that $\hat{{\bf b}}=\hat{{\bf z}}$, the Vlasov
equation has the form
\[
\frac{\partial n\bfrt}{\partial t}
 =
 \frac{\partial n\bfrt}{\partial x}
 \frac{\partial \Phi\bfrt}{\partial y}
 -
 \frac{\partial n\bfrt}{\partial y}
 \frac{\partial \Phi\bfrt}{\partial x}
 =
 -\left[ n\bfrt, h\bfrt\right],
\]
where the single-guiding-center Hamiltonian is
\[
 h\bfrt = -\Phi\bfrt,
\]
and we have defined the corresponding Poisson bracket
\[
\left[ a\bfrt,b\bfrt \right]
 =
 \frac{\partial a\bfrt}{\partial x}
 \frac{\partial b\bfrt}{\partial y}
 -
 \frac{\partial a\bfrt}{\partial y}
 \frac{\partial b\bfrt}{\partial x},
\]
so that $x$ and $y$ are canonically conjugate variables.  The
configuration space can thus be regarded as a phase space, and the
dynamics of the plasma are then a symplectomorphism in configuration
space.

The corresponding Hamiltonian field structure is noncanonical, and
Lie-Poisson in form~\cite{bib:morrison}.  That is, the equation of
motion is
\[
\frac{\partial n\bfrt}{\partial t}
 =
 \left\{ n\bfrt,H[n] \right\},
\]
where the field Lie-Poisson bracket of two functionals of $n$ is
\[
\left\{ A,B \right\}
 =
 \int\dtr n\bfrt
 \left[
 \frac{\delta A}{\delta n\bfrt},
 \frac{\delta B}{\delta n\bfrt}
 \right],
\]
and the field Hamiltonian functional is given by
\bgea
 H[n] &=&  \frac{1}{2}\int\dtr n\bfrt h\bfrt
       =  -\frac{1}{2}\int\dtr n\bfrt\Phi\bfrt \nonumber \\
      &=& -\frac{1}{2}
        \int\dtr n\bfrt
        \int\dtrp n(\bfr',t)
        \grn,
 \label{eq:fham}
\eea
where, in turn, $\grn$ denotes the Green's function of the Poisson
problem,
\[
 \laplacian\grn = 4\pi\dfunr.
\]
The factor of $\smallhalf$ in the Hamiltonian prevents double counting
of the energy.

The Lie-Poisson bracket admits the infinite set of Casimir
functionals~\cite{bib:morrison},
\[
 {\cal Z}_\psi [n] \equiv \int\dtr\psi\left(n\bfrt\right),
\]
where $\psi$ is any function of its argument.  These Casimir functionals
commute with any other functional, including the Hamiltonian, and hence
they are constants of the motion.  We may span the set of Casimir
functionals with analytic $\psi$ by the set
\[
 {\cal Z}_j [n] \equiv \frac{1}{j}\int\dtr n^j\bfrt,
\]
indexed by the integers $j\geq 1$.  The Casimir functional ${\cal Z}_2$
is of special importance; it is called the {\it enstrophy}, since its
analog for the Euler fluid is the integral of the square of the
vorticity.

If we further suppose that the Penning trap is cylindrically symmetric,
with a grounded outer wall, then the Hamiltonian is invariant under
rotation and time translation, so that the angular momentum,
\[
L[n] = \int\dtr r^2 n\bfrt,
\]
and the total energy, $H[n]$, are also good invariants.

\subsection{Variational Descriptions of the Metaequilibrium State}
\label{ss:vara}

In spite of the elegance of this Hamiltonian structure, both laboratory
and numerical experiments indicate that some of these theoretical
invariants are broken, presumably due to collisional effects that are,
of course, ignored in a Hamiltonian formulation.

Stable equilibria, both axisymmetric and nonaxisymmetric, have been
observed for this system~\cite{bib:nonax}.  If the plasma is initialized
with a hollow density profile, however, the spatial gradients will
excite diochotron (Kelvin-Helmholtz-like) instabilities on a short time
scale, which will, in turn, give rise to much longer-lived vortex
patches.  As these patches move about and collide, they shed filiaments
of particles which erode the vortex patches further, until a {\it
metaequilibrium} state with a characteristic profile shape is eventually
reached.  This metaequilibrium state can persist for tens of thousands
of diochotron periods, until it is finally destroyed by viscous effects
which are outside the scope of this paper~\cite{bib:iso,bib:huang1}.
Here we focus on the metaequilibria of initially axisymmetric
configurations.

In the course of the above-described evolution, the total mass ${\cal
Z}_1$ and the angular momentum $L$ are well conserved.  The energy $H$
is reasonably well conserved.  The enstrophy ${\cal Z}_2$ tends to
decrease in more-or-less monotonic fashion, and other Casimir
invariants, such as the Boltzmann entropy
\[
 S[n] = -\int\dtr n\bfrr\ln\left[n\bfrr\right],
\]
are badly broken.  For this reason, ${\cal Z}_1$, $L$, and $H$, are
often referred to as {\it robust} or {\it rugged} invariants, while the
${\cal Z}_j$ with $j\geq 2$ are termed {\it fragile} or {\it dissipated}
invariants~\cite{bib:huang1}.

It is tempting to try to derive the shape of the final profile from a
variational principle.  Most work has centered on maximizing the
Boltzmann entropy, under the condition that the robust invariants are
fixed~\cite{bib:maxent}.  Using the Boltzmann entropy, one can demand
\[
 0 = \delta (S - \alpha {\cal Z}_1 - \beta H - \lambda L),
\]
which yields the relationship
\bge
 -1-\ln\left[ n\bfrr\right] + \beta \Phi\bfrr =
 \alpha + \lambda r^2.
 \label{eq:maxbent}
\ee
Taking the Laplacian of both sides, we arrive at an equation for the
density profile,
\bge
 -\laplacian\left[\ln n\bfrr\right] + 4\pi\beta n\bfrr =
 4\lambda.
 \label{eq:profa}
\ee
Unfortunately, the observed metaequilibrium density profiles are
significantly more peaked than the solutions to this
equation~\cite{bib:huang1,bib:huang2}.

Matthaeus and Montgomery~\cite{bib:mm} have suggested that turbulent
relaxation follows a {\it Selective Decay Hypothesis}: The approach to
equilibrium is governed by the most slowly decaying fragile invariant.
In this case, because the enstrophy seems to be the most slowly decaying
of all the fragile invariants, it has been proposed that nonneutral
plasmas~\cite{bib:huang2} and Euler turbulence~\cite{bib:leith} tend to
minimize enstrophy, rather than maximize the Boltzmann entropy, while
still respecting the robust invariants.  Indeed, if we replace the above
variational principle with
\[
 0 = \delta ({\cal Z}_2 - \alpha {\cal Z}_1 - \beta H - \lambda L),
\]
then we are quickly led to the relationship,
\bge
 n\bfrr + \beta \Phi\bfrr = \alpha + \lambda r^2.
 \label{eq:minenst}
\ee
Taking the Laplacian of both sides, we arrive at the linear Helmholtz
equation for the density profile,
\bge
 \laplacian n\bfrr + 4\pi\beta n\bfrr = 4\lambda.
 \label{eq:profb}
\ee
The well behaved cylindrically symmetric solutions to this equation are
of the form
\bge
 n(r) = \nu J_0(\kappa r) + \mu,
 \label{eq:forma}
\ee
where $J_0$ is the Bessel function, and the constants $\mu$, $\nu$, and
$\kappa$, which have replaced $\beta$, $\lambda$, and the constant of
integration, are fully determined by the constrained quantities, ${\cal
Z}_1$, $H$, and $L$.

Because the solutions to the above variational problem often predict a
negative density near the wall, Huang and Driscoll also introduced a
{\it Restricted Minimum Enstrophy} (RME) model~\cite{bib:huang2} in
which the above profile is replaced by the cutoff form
\bge
 n(r) =
  \left\{
   \begin{array}{ll}
    \nu\left[ J_0 (\kappa r) - J_0 (\kappa r_0)\right] &
    \mbox{for $0\leq r\leq r_0$} \\
    0 &
    \mbox{for $r_0\leq r\leq r_w$}
   \end{array}
  \right. ,
 \label{eq:profc}
\ee
where $r_w$ is the wall radius (which, in dimensionless units, can be
set to unity).  The constant $r_0$ has replaced the constant $\mu$; all
three constants are still fully determined by the three constraints.
This form is justified only by the observation that a negative density
near the wall cannot be physical, and that nonmonotonic profiles are
typically subject to diochotron instabilities.

Huang and Driscoll then carefully compared~\cite{bib:huang2}
experimental data with the profiles generated by Eqs.~(\ref{eq:profa}),
(\ref{eq:profb}), and (\ref{eq:profc}).  They found that the data
clearly ruled out the maximum Boltzmann entropy profile of
Eq.~(\ref{eq:profa}).  The minimum enstrophy profile of
Eq.~(\ref{eq:profb}) was much better.  Best of all was the RME profile
of Eq.~(\ref{eq:profc}).  The experimental data was clearly consistent
with the Bessel function profiles.

To date, a completely satisfactory explanation of this tendency to
minimize enstrophy, rather than maximize entropy, does not exist.  In
the remainder of this paper, we shall show that this phenomenon is
consistent with a generalization of thermodynamics and statistical
physics recently proposed by Tsallis~\cite{bib:tsallis1}.  Though this
is still not an explanation per se, it certainly makes possible the
association of this phenomenon with a much larger -- and growing -- body
of research.

\section{Generalized Thermodynamics}
\label{s:genthermo}

Tsallis~\cite{bib:tsallis1} has proposed a generalization of
thermodynamics and statistical physics to describe systems with
long-range interactions, or with long-time memory.  For a system with
$W$ microscopic state probabilities $p_i\geq 0$, that are normalized
according to
\bge
 1 = \sum_i^W p_i,
 \label{eq:norm}
\ee
Tsallis bases his formalism upon the following two axioms:
\begin{axiom}
The entropy of the system is given by
\[
 S_q = k \frac{1-\sum_i^W p_i^q}{q-1}
     = \frac{k}{q-1}\sum_i^W \left( p_i - p_i^q \right),
\]
where $k$ and $q$ are real constants.
\end{axiom}
\begin{axiom}
An experimental measurement of an observable $O$, whose value in state
$i$ is $o_i$, yields the $q$-expectation value,
\[
 O_q = \sum_i^W p_i^q o_i,
\]
\end{axiom}
of the observable $O$.

It is to be emphasized that these statements are taken as {\it axioms}.
As such, their validity is to be decided solely by the conclusions to
which they lead, and ultimately by comparison with experiment.

We first note that in the limit as $q$ approaches unity we recover the
familiar expressions
\[
 S_1 = -k \sum_i p_i \ln p_i
\]
and
\[
 O_1 = \sum_i p_i o_i,
\]
whence we may identify $k$ with Boltzmann's constant, $k_B$.  More
generally, it has been noted~\cite{bib:tsallis2} that $k$ may be
$q$-dependent, and need only coincide with Boltzmann's constant for
$q=1$; for the purposes of this paper, however, we disregard that
possibility and henceforth adopt units so that $k=k_B=1$.  In any case,
it is clear that Tsallis' thermodynamics contain the more orthodox
variety as a special case.

The success of thermodynamics and statistical physics depends crucially
upon certain properties of the entropy and energy, and much effort has
been devoted to showing that many of these are valid for arbitrary $q$,
and to finding appropriate generalizations of the rest.  Following
Tsallis' presentation~\cite{bib:tsallis3}, it is straightforward to
verify the following properties:
\begin{property}
The generalized entropy is positive.
\end{property}
That is, we have $S_q \geq 0$, where equality holds for pure states
($\exists i : p_i=1$) and for $q > 0$.
\begin{property}
The microcanonical ensemble has equiprobability.
\end{property}
To see this, we extremize the generalized entropy under the constraint
of normalized probabilities, Eq.~(\ref{eq:norm}).  Introducing the
Lagrange multiplier $\lambda$, we set
\[
 0 = \frac{\partial}{\partial p_i}
     \left(S_q - \lambda\sum_i^W p_i\right)
   = -\frac{q}{q-1} p_i^{q-1} - \lambda.
\]
It follows that
\[
 p_i = \left[\frac{\lambda (1-q)}{q}\right]^\frac{1}{q-1}.
\]
Since this is independent of $i$, imposition of the constrant,
Eq.~(\ref{eq:norm}), immediately yields $p_i=1/W$.
\begin{property}
The entropy is concave for $q > 0$ and convex for $q < 0$.
\end{property}
This follows immediately from the Hessian matrix,
\[
 \frac{\partial^2}{\partial p_i \partial p_j}
 \left(S_q - \lambda\sum_i^W p_i\right) =
 -q p_i^{q-2}\delta_{ij},
\]
which is clearly negative (positive) definite for $q > 0$ ($q < 0$).
Thus, the generalized entropy is maximized for $q > 0$, and minimized
for $q < 0$.

Next, we consider the canonical ensemble.  If we define a state energy
$\varepsilon_i$, so that the generalized internal energy is given by
\[
 U_q = \sum_i^W p_i^q \varepsilon_i,
\]
then we can extremize $S_q$ under the constraint that probability is
conserved and that the energy is fixed.  We find
\begin{property}
The canonical ensemble probability distribution is
\[
 p_i = \frac{1}{Z_q}
       \left[
        1 - (1-q)\beta\varepsilon_i
       \right]^\frac{1}{1-q},
\]
where we have defined the generalized partition function
\[
 Z_q \equiv
  \sum_i^W
       \left[
        1 - (1-q)\beta\varepsilon_i
       \right]^\frac{1}{1-q},
\]
and where we have defined the inverse temperature, $\beta\equiv 1/T$.
\end{property}
We note that, in the limit as $q$ approaches unity, we recover the
familiar expressions
\[
 p_i = \frac{e^{-\beta\varepsilon_i}}{Z_1}
\]
and
\[
 Z_1 \equiv
  \sum_i^W e^{-\beta\varepsilon_i}.
\]

For $q\neq 1$, we note that the absolute value of the energy matters --
an additive constant in the energy spectrum will produce physical
effects.  Moreover, we note that, for generic real values of $q$, the
above expression for $p_i$ breaks down if $1-(1-q)\beta\varepsilon_i <
0$.  In such a situation, state $i$ is {\it thermally forbidden}.  For a
positive energy spectrum that is unbounded above, and assuming that
$\beta > 0$, this will happen for sufficiently high $\varepsilon_i$ if
$q < 1$.  Thus, in this situation, the Tsallis distribution has a
natural cutoff in energy for $q < 1$.

More significantly, we note that
\begin{property}
The Legendre-transform structure of thermodynamics is invariant for all
$q$.
\end{property}
To see this, we first note that
\[
 -\frac{\partial}{\partial\beta}
 \left(
  \frac{Z_q^{1-q} - 1}{1-q}
 \right) = U_q,
\]
whence we identify the free energy
\[
F_q = -\frac{1}{\beta}
      \left(
       \frac{Z_q^{1-q}-1}{1-q}
      \right).
\]
It is then possible to verify that
\[
 F_q = U_q - T S_q,
\]
and it follows that
\[
 \frac{\partial S_q}{\partial U_q} = \frac{1}{T}.
\]
As Tsallis points out~\cite{bib:tsallis4}, these equations lie at the
very heart of thermodynamics, and the fact that they are invariant under
$q$ is significant.  The grand canonical ensemble has also been
treated~\cite{bib:tsallis5}.

The most striking and significant differences between the generalized
thermodynamics and the more usual variety have to do with the
extensivity of the state variables.  If we partition the microscopic
states of the system into two disjoint subsets, $L=\{ 1,\ldots,V\}$ and
$R=\{ V+1,\ldots,W\}$, with respective probabilities
\[
 p_L \equiv \sum_{i=1}^V p_i
\]
and
\[
 p_R \equiv \sum_{i=V+1}^W p_i,
\]
then it is straightforward to verify that
\begin{property}
The generalized entropy obeys the following generalization of the
Shannon Additivity Property,
\[
 S_q(p_1,\ldots, p_W) =
 p_L^q
 S_q
  \left(
  \frac{p_1}{p_L},
  \ldots,
  \frac{p_V}{p_L}
  \right) +
 p_R^q
 S_q
  \left(
  \frac{p_{V+1}}{p_R},
  \ldots,
  \frac{p_W}{p_R}
  \right) +
 S_q
  (p_L,p_R).
\]
\end{property}
Alternatively, we can consider the total entropy of two completely
independent subsystems, $A$ and $B$.  Since the subsystems are
independent, the probability that their union $A\cup B$ has subsystem
$A$ in state $i$ and subsystem $B$ in state $j$ is given by
\[
 p^{A\cup B}_{ij} = p^A_i p^B_j.
\]
After a bit of algebra, we find that
\begin{property}
The generalized entropy obeys the following additivity rule
\[
 S_q(A\cup B) = S_q(A) + S_q(B) + (1-q) S_q(A) S_q(B),
\]
and is thus superadditive (entropy of whole is greater than the sum of
its parts) for $q < 1$ and subadditive for $q > 1$.
\end{property}
Likewise, we find
\begin{property}
The generalized expectation value of an observable $O$ obeys the
following additivity rule
\[
 O_q(A\cup B) = O_q(A) + O_q(B) +
 (1-q)\left[O_q(A) S_q(B) + O_q(B) S_q(A)\right].
\]
\end{property}
Note that, in both cases, extensivity is recovered only when $q=1$.

It is believed -- but not proven at the time of this writing -- that the
Tsallis entropy is the only one for which all of the above properties
hold.  Moreover, generalized versions of the Boltzmann
H-theorem~\cite{bib:hth}, fluctuation-dissipation
theorem~\cite{bib:fdt}, and Onsager reciprocity theorem~\cite{bib:ort}
exist for all $q$.  The formalism is thus an important generalization of
most of the principal results of thermodynamics and statistical physics.

Of course, to verify that this generalization is useful, it is necessary
to show that it holds for certain physical systems with values of $q$
that are different from unity.  In the past two years, much work has
been done along these lines, and the method has been applied with great
benefit to astrophysical problems such as stellar
polytropes~\cite{bib:pp1}, L\'{e}vy flights~\cite{bib:levy}, the
specific heat of the hydrogen atom~\cite{bib:specheat}, and numerous
other physical systems~\cite{bib:bio,bib:simanneal}.  For some of these
systems, strict inequalities have been proven, demonstrating that $q$
must be different from unity in order to obtain a consistent
thermodynamic description~\cite{bib:pp1}.

\section{Stellar Polytropes}
\label{s:stellar}

\subsection{Hydrostatic Equilibria}

One of the first problems to which Tsallis' thermodynamics was
applied~\cite{bib:pp1} was that of stellar polytropes, first studied by
Lord Kelvin~\cite{bib:kelvin}, and treated in detail by
Chandrasekhar~\cite{bib:chand}.  Because stellar polytropes are
equilibria of the Poisson-Vlasov equations, they are highly relevant to
the present study.  Therefore, in this section, we shall review the
previous application of Tsallis' formalism to this problem by Plastino
and Plastino~\cite{bib:pp1}.  We note that they used a energy functional
that was linear in the distribution function, whereas the full
Poisson-Vlasov energy functional is quadratic.  In this treatment, we
use the full quadratic functional, and compare our analysis to theirs.
One of the side benefits of this treatment is that it shows the
extension of Tsallis' second axiom to observables that are quadratic
functionals of the distribution.

A polytropic process has the equation of state
\[
 P = K\rho^\gamma,
\]
where $P$ is the pressure, $\rho$ is the density, and $\gamma$ is a
constant that can be related to the specific heats.  If $\Phi\bfrr$
denotes the gravitational potential, then the hydrostatic equilibrium is
given by
\[
 0 = -\grad P - \rho\grad\Phi.
\]
It follows that
\[
 0 = \grad\left(\rho^{\gamma - 1}+\frac{\gamma - 1}{K\gamma}\Phi\right).
\]
We now seek solutions with compact support in domain ${\cal D}$.  If we
require that the density $\rho$ vanish on the boundary $\partial {\cal
D}$, then we must have the following relationship between $\rho$ and
$\Phi$,
\[
 \rho =
 \left[
 \frac{\gamma - 1}{K\gamma}\left(\Phi^{(0)} - \Phi\right)
 \right]^{\frac{1}{\gamma-1}},
\]
where $\Phi^{(0)}$ is the potential on the boundary.  The nonlinear
Poisson equation for the gravitational potential is then
\bge
 \laplacian\Psi = -C\Psi^{\frac{1}{\gamma-1}},
 \label{eq:sp}
\ee
where $\Psi\equiv\Phi^{(0)}-\Phi$, and $C$ is a constant.  The boundary
condition is that $\Psi=0$ on $\partial {\cal D}$.  This equation has,
for example, spherically symmetric solutions, corresponding to compact
spherical configurations of self-gravitating mass, that are called {\it
stellar polytropes}.

\subsection{Kinetic Equilibria}

As an alternative to the above hydrodynamic description, we can seek
polytropic equilibria of the Vlasov equation for the mass distribution
function $f(z, t)$ where $z=(\bfr,\bfv)$ coordinatizes the phase space
of the system.  As is well known, the equilibria of the Vlasov equations
are functions of the constants of the motion.  We denote the (negative
of the) total energy by
\[
 {\cal E}(z) \equiv \Psi\bfrr - \frac{m}{2}v^2,
\]
so that a marginally confined particle on $\partial {\cal D}$ with zero
velocity has ${\cal E}=0$, and a confined particle has ${\cal E} > 0$.
Noting that any function of ${\cal E}$ is a solution of the Vlasov
equation, we examine solutions of the form
\[
 f =
  \left\{
   \begin{array}{ll}
    \theta {\cal E}^{n-3/2} & \mbox{for ${\cal E} > 0$} \\
    0                       & \mbox{for ${\cal E} \leq 0$}
   \end{array}
  \right. ,
\]
where $\theta$ is a constant.  The mass density of a spherically
symmetric configuration is then given by
\bge
 \rho\bfrr =
   \int dz'\; f(z') \dfunr =
   \int d^3v'\; f(\bfr, \bfv') =
   \theta\int_0^{\sqrt{2\Psi}} dv\; 4\pi v^2
   \left[\Psi\bfrr - \frac{v^2}{2}\right]^{n-3/2}.
 \label{eq:rho}
\ee
The integral gives rise to a beta function which can be expressed in
terms of gamma functions to finally yield
\bge
 \rho\bfrr =
  (2\pi)^{3/2}\theta\;
  \frac{\Gamma (n-\smallhalf)}{\Gamma (n+1)}
  \Psi^n\bfrr.
 \label{eq:gamma}
\ee
Comparing this to Eq.~(\ref{eq:sp}), we can identify
\[
 n = \frac{1}{\gamma - 1}
\]
or
\[
 \gamma = 1 + \frac{1}{n}.
\]

\subsection{Variational Principle with Linear Energy Functional}
\label{ss:lin}

Note that the stellar polytropes comprise a one-parameter family of
equilibria, where the parameter is $\gamma$ (or, equivalently, $n$).  We
now examine the question of whether or not these polytropic equilibria
are thermodynamically stable, in the sense that they can be obtained
{}from an entropic variational principle, and, if so, for what values of
the parameter $\gamma$ (or $n$) this is possible.

Plastino and Plastino have addressed this question~\cite{bib:pp1} by
extremizing Tsallis' entropy for this problem
\[
 S_q[f] = \frac{1}{q-1} \int dz\; \left[f(z) - f^q(z)\right],
\]
under the constraints of fixed mass and energy expectation values,
\[
 M_q[f] = \int dz\; f^q(z)
\]
\bge
 U_q[f] = \int dz\; f^q(z)
  \left(\frac{v^2}{2} + \Phi\bfrr\right).
 \label{eq:lenergy}
\ee
In fact, they used $M_1$ and $U_1$ in their work, because this was
before Tsallis had advanced his second axiom about expectation values.
This issue was subsequently rectified in a paper by
Tsallis~\cite{bib:tsallis6}, and we present only the corrected version
here.

Introducing Lagrange multipliers, the variational problem
\[
 0 = \delta\left( S_q + \alpha M_q + \beta U_q \right)
\]
yields the equilibrium distribution
\bgea
 f(z)
  &=&
  \left\{
   q
   \left[
    1 -
    (q-1) \alpha -
    (q-1) \beta \left(\frac{v^2}{2} + \Phi\bfrr\right)
   \right]
  \right\}^{\frac{1}{1-q}} \nonumber \\
  &=&
  \left\{
   q
   \left[
    1 -
    (q-1) \left(\alpha - \Phi^{(0)}\right) +
    (q-1) \beta {\cal E}(z)
   \right]
  \right\}^{\frac{1}{1-q}}.
  \label{eq:eqdista}
\eea
For this to be a power law in ${\cal E}$, we select $\alpha$ so that
$1-(q-1)(\alpha-\Phi^{(0)})=0$, so we can write $f(z) = D {\cal
E}^{1/(1-q)}(z)$, where $D$ is a constant.  The density measured at a
point $\bfr$ is then given by the $q$-expectation value of the spatial
delta function,
\[
 \rho_q\bfrr
  = \int dz'\; f^q(z')\dfunr
  = D^q \int d^3v'\; {\cal E}^{\frac{q}{1-q}}(\bfr, \bfv').
\]
We see that this corresponds to Eq.~(\ref{eq:rho}) if we identify
$n-\threehalves=q/(1-q)$, or
\[
 n = \frac{3}{2} + \frac{q}{1-q}.
\]
As pointed out by Plastino and Plastino~\cite{bib:pp1}, it is known that
$n$ must exceed $1/2$ in order to avoid the singularity in the gamma
function in Eq.~(\ref{eq:gamma}), but that values in excess of $5$ give
rise to unnormalizable mass distributions, and are therefore unphysical.
This means that $q\in (-\infty,\sevenninths)$.  Thus, stellar polytropes
cannot be described thermodynamically unless $q$ values less than
$\sevenninths$ are used.

Finally, note that the aforementioned cutoff of the distribution with
energy -- a generic feature of Tsallis distributions with $q<1$ --
naturally gives rise to the spatial cutoff of the mass distribution, and
hence the compact nature of the stellar polytrope.

\subsection{Variational Principle with Quadratic Energy Functional}

Note that the energy functional in Eq.~(\ref{eq:lenergy}) was regarded
as linear in $f^q(z)$ in the above analysis.  Specifically, in deriving
Eq.~(\ref{eq:eqdista}), we wrote the functional derivative of $U_q$ with
respect to $f(z)$ as
\[
 \frac{\delta U_q}{\delta f(z)}
 =
 qf^{q-1}(z)\left[\frac{v^2}{2}+\Phi\bfrr\right].
\]
Strictly speaking, this is not correct because the potential $\Phi\bfrr$
depends on $f(z)$, and we did not account for this in the above
variation.  This is precisely the problem of self-consistency of the
field -- a crucial feature of the Poisson-Vlasov
system~\cite{bib:morrison}.  To correct this problem, it is best to
write the energy as a quadratic functional of the distribution, just as
in Eq.~(\ref{eq:fham}) -- only now $q$-expectation values should be used
throughout.  Thus,
\[
 U^Q_q[f] =
  \int dz\; f^q(z)\frac{v^2}{2} +
  \frac{1}{2} \int dz\; f^q(z) \int dz'\; f^q(z')\grn,
\]
where the superscript $Q$ denotes {\it quadratic}, the factor of
$\smallhalf$ in front of the potential prevents double-counting of the
energy, and $\grn$ is the Green's function for Poisson's equation which
satisfies
\[
 \laplacian\grn = 4\pi\dfunr.
\]
Note that the functional derivative of $U^Q_q$ with respect to $f(z)$ is
now
\[
\frac{\delta U^Q_q}{\delta f(z)}
 =
 qf^{q-1}(z)
 \left[
  \frac{v^2}{2} + \Phi_q\bfrr
 \right],
\]
where we have defined
\[
 \Phi_q\bfrr\equiv\int dz'\; f^q(z')\grn,
\]
which in turn satisfies
\bge
 \laplacian \Phi_q\bfrr = 4\pi \rho_q\bfrr.
 \label{eq:poissonq}
\ee

The variational principle thus results in an equation very similar to
Eq.~(\ref{eq:eqdista}), except with ${\cal E}$ replaced by
\[
 {\cal E}_q\equiv\Psi_q\bfrr-\frac{v^2}{2},
\]
where in turn
\[
 \Psi_q\bfrr\equiv\Phi^{(0)}_q-\Phi_q\bfrr.
\]
The resulting expression for $\rho$ is then a power law in $\Psi_q$,
rather than in $\Psi$.  We still conclude that
\[
 n = \frac{3}{2} + \frac{q}{1-q},
\]
and so the upper bound on $q$ of $\sevenninths$ still holds.  Note that
the nonlinear Poisson equation for the gravitational potential,
\[
 \laplacian\Psi_q = -C\Psi_q^n,
\]
is now satisfied by $\Psi_q$, rather than by $\Psi_1$.

This exercise might be dismissed as demonstrating little more than the
fact that the potential used in Subsection~\ref{ss:lin} should be
interpreted as the $q$-expectation value of the Green's function, rather
than as the usual one.  This objection notwithstanding, the derivation
using the quadratic energy functional has the following virtues:
\begin{itemize}
\item It more clearly shows that the conclusions reached by this method
are valid for the system of particles in their own {\it self-consistent}
field.
\item It demonstrates that Tsallis' second axiom extends in a natural
way to quadratic functionals of distributions.
\item It yields the natural generalization of the potential and the
density, and shows that the form of Poisson's equation,
Eq.~(\ref{eq:poissonq}), relating them is $q$-invariant.
\item It is generally more consistent with the flavor and spirit of
Tsallis' formalism than previous derivations.
\end{itemize}

\section{Generalized Thermodynamic Description of Euler Turbulence}
\label{s:turb}

We now return to the problem of deriving the metaequilibrium profiles of
relaxed Euler turbulence.  We redo the calculation of
Subsection~\ref{ss:vara}, using the Tsallis prescriptions for the
entropy and the robustly conserved quantities.  The entropy is thus
\[
 S_q[n] =
  \frac{1}{q-1}
  \int\dtr \left[ n\bfrr - n^q\bfrr\right],
\]
and the constraints, expressed in terms of $q$-expectation values, are
\[
 {\cal Z}_q[n] =
  \frac{1}{q}
  \int\dtr n^q\bfrr
\]
\[
 H_q[n] =
  -\frac{1}{2}\int\dtr n^q\bfrr \int\dtrp n^q\bfrrp \grn
\]
and
\[
 L_q[n] =
  \int\dtr r^2 n^q\bfrr.
\]
Setting
\[
 \delta
  \left(
   S_q - \alpha {\cal Z}_q - \beta H_q - \lambda L_q
  \right) = 0,
\]
we find
\bge
 \frac{n^{1-q}\bfrr-q}{q(q-1)} +
 \beta\Phi_q\bfrr = \alpha + \lambda r^2,
 \label{eq:maxtent}
\ee
where
\[
 \Phi_q\bfrr\equiv
  \int\dtrp n^q\bfrrp\grn
\]
satisfies
\bge
 \laplacian\Phi_q\bfrr = 4\pi n^q\bfrr.
 \label{eq:nlap}
\ee
Applying the Laplacian to Eq.~(\ref{eq:maxtent}), we obtain the
nonlinear Helmholtz equation,
\bge
 \frac{1}{q(q-1)}\laplacian\left[n^{1-q}\bfrr\right] +
  4\pi\beta n^q\bfrr = 4\lambda.
 \label{eq:nhe}
\ee

Now, the {\it observed} particle density $\rho_q\bfrr$ -- i.e., that
which is measured in any experiment -- is then the $q$-expectation value
of the delta function,
\[
 \rho_q\bfrr = \int\dtrp n^q\bfrrp\dfunr = n^q\bfrr.
\]
In terms of this, Eqs.~(\ref{eq:maxtent}), (\ref{eq:nlap}), and
(\ref{eq:nhe}) can be written
\[
 \frac{\rho_q^\frac{1-q}{q}\bfrr-q}{q(q-1)} +
 \beta\Phi_q\bfrr = \alpha + \lambda r^2,
\]
\[
 \laplacian\Phi_q\bfrr=4\pi\rho_q\bfrr
\]
\[
 \frac{1}{q(q-1)}\laplacian\left[\rho_q^\frac{1-q}{q}\bfrr\right] +
  4\pi\beta\rho_q\bfrr = 4\lambda.
\]
As $q\rightarrow 1$, it is seen that this reproduces the results of the
maximum Boltzmann entropy relationship, Eqs.~(\ref{eq:maxbent}) and
(\ref{eq:profa}).  When $q=\smallhalf$, on the other hand, we see that,
within trivial redefinitions of the Lagrange multipliers and the use of
$\Phi_{1/2}$ instead of $\Phi$, this reproduces the results obtained by
minimizing the enstrophy, Eqs.~(\ref{eq:minenst}) and (\ref{eq:profb}),
{\it but for a completely different reason}.  Moreover, just as in the
example of the stellar polytrope, the cutoff in density at a finite
radius $r_0$ appears as a {\it completely natural and generic feature of
the Tsallis distribution, since $q<1$, and does not need ad hoc
justification}.  Thus, we conclude that all prior observations that have
indicated that the relaxation of two-dimensional Euler turbulence tends
to follow the RME principle of Huang and Driscoll can now be
reinterpreted as rather indicating that it maximizes the Tsallis entropy
$S_q$ for $q=\smallhalf$.

We note in passing that there is an easier way to obtain the result that
$q=\smallhalf$.  Without going through this analysis, we note that the
enstrophy itself looks rather like (a linear function of) the Tsallis
entropy with $q=2$.  Of course, this is misleading, because it is
necessary to use $q$-expectation values in the extremization process.
Nevertheless, Tsallis~\cite{bib:tsallis6} has shown that one result of
{\it not} using $q$-expectation values in the extremization process is
to effect the transformation $q\rightarrow 1/q$.  Hence, we are again
led immediately to the result $q=\smallhalf$ for this system.

\section{Conclusions}

The tendency of a two-dimensional Euler fluid to minimize enstrophy --
rather than maximize the Boltzmann entropy -- during turbulent
relaxation to a metaequilibrium state has resisted theoretical
explanation to date.  In this work, we have shown that density profiles
resulting from the Restricted Minimum Enstrophy (RME) theory of Huang
and Driscoll also maximize the Tsallis entropy.  We have thereby
provided an alternative way to understand this phenomenon -- one in
which the density cutoff at finite radius emerges quite naturally -- and
to build a consistent thermodynamical and statistical physical
explanation for it.  In the course of doing this, we have verified
Plastino and Plastino's upper bound of $\sevenninths$ on $q$ for the
stellar polytrope problem using the full quadratic energy functional for
the Poisson-Vlasov system; we have also demonstrated the $q$-invariance
of the Poisson equation for these systems.

While still short of a first-principles {\it explanation} of the RME
model -- it would be nice, for example, to have an a priori way of
knowing {\it why} $q$ should be equal to $\smallhalf$ for this system --
the observation that RME is consistent with Tsallis statistics does
effectively associate it with a large and rapidly expanding body of
theory.  In recent years, generalized thermodynamics has been used to
describe numerous, widely disparate physical systems, with long-range
interactions, long-time (non-Markovian) memory, or fractal space-time
structure, that have resisted previous attempts at a thermodynamic
description.  It is hoped that this observation will stimulate further
research in the use of generalized thermodynamics to describe fluid
turbulence.

\section*{Acknowledgements}

The author would like to thank Constantino Tsallis, Jonathan Wurtele,
and Joel Fajans for their careful review of the draft paper, and their
many helpful suggestions.  This work was supported in part by the Air
Force Office of Scientific Research and the Department of Energy,
Division of High Energy Physics.

\end{document}